\documentclass[twocolumn,groupedaddress,
superscriptaddress,preprintnumbers,
amsmath,amssymb,
aps,
]{revtex4}
\usepackage{graphicx}
\usepackage{dcolumn}
\usepackage{bm}%
\usepackage[colorlinks=true,citecolor=blue]{hyperref}
\hypersetup{colorlinks=true,citecolor=blue,linkcolor=red,urlcolor=blue}

\usepackage{float}
\usepackage{amsmath}
\newcommand{\angstrom}{\mbox{\normalfont\AA}}
\usepackage{color}

\makeatletter
\def\@bibdataout@aps{%
\immediate\write\@bibdataout{%
@CONTROL{%
apsrev41Control%
\longbibliography@sw{%
    ,author="08",editor="1",pages="1",title="0",year="1"%
    }{%
    ,author="08",editor="1",pages="1",title="",year="1"%
    }%
  }%
}%
\if@filesw \immediate \write \@auxout {\string \citation {apsrev41Control}}\fi 
}
\makeatother

\begin{document}

\title{ Charge density wave and superconducting phase in monolayer InSe}
\author{Mohammad Alidoosti}
\affiliation{School of Nano Science, Institute for Research in Fundamental Sciences (IPM), Tehran 19395-5531, Iran}
\author{Davoud  Nasr Esfahani}
\affiliation{Pasargas Institute for Advanced Innovative Solutions(PIAIS), Tehran, Iran}
\affiliation{Department of Converging Technologies, Khatam University, Tehran, Iran}
\author{Reza Asgari}
\email{asgari@ipm.ir}
\affiliation{School of Physics, Institute for Research in Fundamental Sciences (IPM), Tehran 19395-5531, Iran}
\affiliation{ARC Centre of Excellence in Future Low-Energy Electronics Technologies, UNSW Node, Sydney 2052, Australia}
\date{\today}

\begin{abstract}
In this paper, the completed investigation of a possible superconducting phase in monolayer indium selenide is determined using first-principles calculations for both the hole and electron doping systems. The hole-doped dependence of the Fermi surface is exclusively fundamental for monolayer InSe. It leads to the extensive modification of the Fermi surface from six separated pockets to two pockets by increasing the hole densities. For low hole doping levels of the system, below the Lifshitz transition point,  superconductive critical temperatures $T_c \sim 55-75$ K are obtained within anisotropic Eliashberg theory depending on varying amounts of the Coulomb potential from 0.2 to 0.1. However, for some hole doping above the Lifshitz transition point, the combination of the temperature dependence of the bare susceptibility and the strong electron-phonon interaction gives rise to a charge density wave that emerged at a temperature far above the corresponding $T_c$. Having included non-adiabatic effects, we could carefully analyze conditions for which either a superconductive or charge density wave phase occurs in the system. In addition, monolayer InSe becomes dynamically stable by including non-adiabatic effects for different carrier concentrations at room temperature.
\end{abstract}

\maketitle


\section{Introduction}
Motivated by the discovery of graphene~\cite{novoselov2004electric}, a two-dimensional (2D) advanced material with spectacular properties, researchers have greatly discovered 2D layered materials; namely, hexagonal boron nitride \cite{kim2011synthesis}, 
transition metal dichalcogenides (such as MoS$_2$ and WS$_2$) \cite{chhowalla2013chemistry}, magnetic 2D crystalline-like monolayer chromium triiodide (CrI$_3$) \cite{huang2017layer} and other elemental 2D semiconductors such as black phosphorus \cite{li2014black} and silicene \cite{vogt2012silicene}, ranging from insulators, semiconductors, metals, magnetics and superconductors. 

In addition, group III-VI semiconductors (M$_2$X$_2$, M = Ga and In and X = S, Se and Te) with sombrero-shaped valence band edges have shown marvelous electrical and optical properties \cite{zolyomi2013band, PhysRevB.89.205416}. The bulk indium selenide (InSe), III-monochalcogenide semiconductor, has $\beta,\, \varepsilon $ and $ \gamma $ structural phases depending on the stacking characteristics \cite{han2014indium, lei2014evolution, politano2017indium}. Among these phases, $\varepsilon$ has an indirect bandgap about $1.4$ eV \cite{lei2014evolution}, while, $\beta$ and $\gamma$ phases have a direct bandgap close to $1.28$ \cite{gurbulak2014structural} and $1.29$ eV \cite{julien2003lithium}, respectively. Electron–phonon coupling (EPC) and the superconductive properties of an electron-doped monolayer InSe were studied \cite{chen2019phonon} and a superconductive transition temperature about 3.41 K was reported.
Moreover, it has been shown that hole states in monolayer InSe are strongly renormalized by coupling with acoustic phonons leading to the formation of satellite quasiparticle states near the Fermi energy~\cite{PhysRevLett.123.176401}.
Not long ago, monolayer InSe has been fabricated from its bulk counterpart by mechanical exfoliation~\cite{mudd2013tuning,mudd2014quantum,feng2015gate}.
High carrier mobility of about $10^{3}$ cm$^{2}$/Vs, which is greater than that of MoS$_2$ \cite{cui2015multi},
has been reported at room temperature \cite{sun2016ab,bandurin2017high}; suggesting that this 2D material is promising for ultra-thin digital electronics applications. Furthermore, InSe represents a promising material for making use of FETs \cite{marin2018first}.

The presence of a sombrero-shaped valence band in the electronic band structure of monolayer InSe
gives rise to a larger density of states (DOS), which is similar to that of one-dimensional material, and specifies a Van Hove singularity at the valence band maximum (VBM) which could primarily lead to a magnetic transition and superconducting phases as well \cite{PhysRev.108.1175,PhysRevLett.114.236602,dresselhaus2007new,hung2017two,wu2014magnetisms}.
Stimulated by the remarkable discovery of gate-induced superconductivity in graphene (upon lithium adsorption) \cite{ludbrook2015evidence,ichinokura2016superconducting,profeta2012phonon,xue2012superconductivity}, a new field for investigating superconducting features on other 2D materials typically has emerged.
In advance, lithium adsorbed graphene was properly utilized for 2D superconductivity. Undoubtedly owing to a small DOS at the Fermi level and $\sigma_h$ symmetry which gives rise to a weakened electron coupling with the flexural modes, graphene illustrates
a small electron-phonon coupling constant, $\lambda$. However, these shortcomings could be lifted by typically making use of lithium adsorption \cite{profeta2012phonon,xue2012superconductivity}.

Even though monolayer InSe naturally has $\sigma_h$ symmetry, electrons in monolayer InSe could couple to the flexural phonons owing to the presence of atomic layers away from the symmetry plane. Notably, this coupling alongside a larger DOS near the VBM potentially leads to a significant EPC parameter. On the other hand, the active presence of a significant DOS as well as $\lambda$ makes the system susceptible to a charge density wave (CDW) instability, which represents a static modulation of the itinerant electrons and usually accompanied by a periodic distortion of the lattice. The CDW formation may naturally arise from a possible combination of a large nesting and/or electron-phonon interaction at a specific phonon wave vector ($\textbf q_{{CDW}}$). Therefore, the formation of the CDW must be carefully examined for systems with a strong EPC, though a superconducting state is possible.

The standard method of properly investigating CDW formation is first to calculate the phonon dispersion of the system within density functional theory (DFT) calculations, i.e. considering either a small displacement or density functional perturbation theory (DFPT) method at specific temperatures \cite{baroni2001phonons}. It is worth mentioning that the long-wavelength electron-phonon interaction induced phonon self-energy is carefully considered in the phonon dispersion of both mentioned approaches \cite{calandra2010adiabatic}. However, it has become evident that dynamical phonons undoubtedly play a significant role and non-adiabatic/dynamic effects could give rise to a significantly renormalized phonon dispersion for doped semiconducting materials \cite{lazzeri2006nonadiabatic,saitta2008giant,novko2019nonadiabatic} including InSe.

Here, we investigate a viable superconducting and CDW phases of monolayer InSe based on DFT and necessary DFPT calculations. We calculate the renormalized phonon dispersion owing to the electron-phonon coupling in both adiabatic and non-adiabatic regimes for different temperatures and doping levels. We further investigate the competition between CDW formation and the superconductive phase for different hole and electron doping levels. We eagerly discuss the most important phonon wave vectors leading to the remarkable electron-phonon coupling strength which well expresses the significance of both bare susceptibility and the nesting function below and above the Lifshitz transition point.
By including non-adiabatic effects, we carefully analyze conditions for which either a superconductive or CDW phase could typically emerge in the system. Our desired results show that in some hole-doped cases, CDW instability prevents access to quite high-temperature superconductivity, whereas for some other doped levels, the achievement of such superconducting temperatures is possible. In the electron-doped cases, the CDW instability is significantly suppressed, and therefore the superconducting phase is possible.

The paper is organized as follows. We commence with a description of our theoretical formalism in Sec. \ref{sec:Theory}, followed by the details of the DFT  and DFPT calculations. Numerical results for the band structures, phonon dispersions, DOS,  superconducting critical temperature, and CDW in adiabatic and non-adiabatic approximations are reported in Sec. \ref{sec:results}. We summarize our main findings in Sec. \ref{sec:conclusion}.


\section{Theory and computational details}\label{sec:Theory}
Self-consistent DFT calculations are carefully performed with LDA-norm-conserving pseudopotential as implemented in the
Quantum Espresso package \cite{0953-8984-21-39-395502}. The phonon dispersion and self-consistent deformed potentials
are calculated based on the DFPT method \cite{baroni2001phonons,wierzbowska2005origins, calandra2010adiabatic}.
The Kohn-sham wave functions and Fourier expansion of the charge density are truncated at 90 and 360 Ry, respectively. To eliminate spurious interactions between adjacent layers, a vacuum space of 25 $\angstrom$ along the $z$ direction is adopted.
For the electronic and phononic calculations, a $24\times24\times1$ $\textbf k$ mesh and $12\times12\times1$ $\textbf q$-mesh,
are used and a finer $\textbf k$ mesh of $240\times240\times1$ and $\textbf q$ mesh of $ 120\times120\times1$, respectively, are applied to calculate the Wannier
interpolation of the electronic and phonon dispersions as implemented in EPW code \cite{marzari1997n,souza2001maximally,mostofi2008wannier90,ponce2016epw}.
The Dirac delta functions are approximated by applying a Gaussian smearing $\sigma_{el} = 5\ \text{meV} $ and $\sigma_{ph} = 0.2 \ \text{meV} $. The convergence of results is carefully performed as a function of the $\textbf k$ and $ \textbf q$ mesh and Gaussian smearing. Moreover, to adequately describe the temperature dependence of the electronic structure, the Fermi-Dirac smearing of about 0.01 Ry is used~\cite{PhysRevB.65.035111}.

Since the static part of the phonon self-energy is typically included in the phonon dispersion, one may uniquely define a dressed phonon frequency as \cite{PhysRevB.95.235121},
\begin{equation}\label{eq1}
\omega^2_{\textbf q , \nu} = \Omega ^2_{\textbf q , \nu} + 2\omega_{\textbf q \nu} \Pi_{\textbf q \nu},
\end{equation}
where $\displaystyle\Omega_{\textbf q , \nu}$ is the bare phonon frequency and
$\Pi_{\textbf q,\nu} = \frac{2}{N_{\textbf k}} \sum_{\textbf k,m,n} |g^{\nu}_{n \textbf k, m\textbf{k+q} }|^2 \frac{f(\varepsilon_{n\textbf{k}})-f(\varepsilon_{m\textbf {k+q}})}{\varepsilon_{n\textbf{k} }- \varepsilon_{m\textbf {k+q}}}$ is the static part of the first-order self-energy of phonon modes, $m \, \text{and}\, n$ refer to the electronic band indeces, $N_{\textbf k}$ is the considerable number of ${\textbf k}$ points, $ g_{mn,\nu}({\bf k,q})$ is the electron-phonon interaction matrix elements and $f (\epsilon)$ represents the Fermi-Dirac distribution function.

It is justifiable to assume $\textbf k$ independent electron-phonon interactions in which $g^{\nu}_{n \textbf k , m \textbf{k+q} }=g^{\nu}_{\textbf{q}n,m}$. Therefore,
Eq.~(\ref{eq1}) can be written as follows;
\begin{equation}\label{eq2}
\omega^2_{\textbf q , \nu} = \Omega ^2_{\textbf q , \nu} + 2\omega_{\textbf q \nu} |g^{\nu}_{\textbf{q}n,m}|^2\chi_0({\textbf q}).
\end{equation}
where $\chi_0(\textbf q) = \frac{2}{N_\textbf{k}} \sum_{\textbf k,m,n} \frac{f(\varepsilon_{n\textbf{k} })-f(\varepsilon_{m\textbf {k+q}})}{\varepsilon_{n\textbf {k} }- \varepsilon_{m\textbf {k+q} }}$
is the bare charge susceptibility. Phonon softening typically emerges at some branches of the phonon spectrum,
known as the Kohn anomaly~\cite{kohn1959image} which originates from any sizable variation of $\chi_0$ as a function of $\bf q$ and/or the electronic temperature. Consequently, it is standard practice to scientifically verify $\chi_0$ as a necessary signature of the phonon softening and thus the formation of the CDW. The CDW instability can be well appeared in the form of an imaginary phonon band when the temperature lies below $T_{CDW}$ (the temperature where softened modes touch the zero frequency at $\textbf q_{CDW}$). 

To estimate the superconducting temperature in systems with a strong EPC, we utilize the Migdal-Eliashberg formalism~\cite{migdal1958interaction,eliashberg1960interactions}
 in the form of a modified Allen-Dynes parametrization~\cite{allen1975transition}:
\begin{equation}
T_{\text{c}} = 
\frac{f_1 f_2 \omega_{\text{log}}}{1.2}
\exp\left(
-\frac{1.04(1+\lambda)}
{\lambda-\mu_{\text{c}}^{*}(1+0.62\lambda)}
\right), 
\label{eq:ADCORR}
\end{equation}	
with $\displaystyle\lambda = 2 \int_{0}^{\infty}\omega^{-1} \alpha^2 \textbf{F}(\omega) d\omega $, $\displaystyle\omega_{\text{log}}=
\exp\left[\frac{2}{\lambda}\int_{0}^{\omega_{\text{max}}} d\omega \frac{\alpha^{2}\textbf{F}( \omega)}{\omega} \log \omega\right]$, 
 $\mu^*_c$ is the Morel-Anderson Coulomb potential, in general, adopted in the range of 0.1-0.2, and $f_1$ and $ f_2$ represent strong-coupling and shape corrections, respectively (for detailed definitions of $f_1$ and 
 $f_2$ see Ref.~\cite{allen1975transition}). 
 The Eliashberg function is defined as
  \begin{eqnarray}
\alpha^{2}\textbf{F}( \omega) = 
\frac{1}{ N(\varepsilon_F)N_\textbf{k} N_\textbf{q}} \sum_{{\bf q,k}\atop \nu , m, n} 
 |g_{mn,\nu}({\bf k,q})|^2\times \\ \nonumber  
\delta(\varepsilon_{n{\bf k}}-\varepsilon_F)\delta(\varepsilon_{m{\bf k}+{\bf q}}-\varepsilon_F) 
\delta (\omega - \omega_{{\bf q}\nu}), 
\label{a2f}
\end{eqnarray}	
where N($\varepsilon_F$) is the electronic density of states at the Fermi level. 
The imaginary part of the phonon self-energy $\gamma_{\textbf q \nu}$ reads as follows:
\begin{equation}
\gamma _{{\bf q}\nu} = \frac{2 
\pi\omega_{{\bf q}\nu}}{N_\textbf{k}}\sum_{mn,{\bf k} }  |g_{mn,\nu}({\bf k,q})|^2  
\delta(\varepsilon_{n{\bf k}}-\varepsilon_F)\delta(\varepsilon_{m{\bf k}+{\bf q}}-\varepsilon_F),
\label{line-width}
\end{equation}	
To carefully analyze the different contributions of $\lambda$ and $\alpha^2\textbf{F}$, the projected quantities are defined as follows. Two principal directions are typically considered: in-plane and out-of-plane distortions. The $\textbf{F}(\omega)$ along the specific direction $\kappa$ is written as

\begin{equation}
\begin{split}
	\textbf{F}^\kappa (\omega)= \sum_{s,\nu}\int \frac{d{\bf q}}{(2\pi)^2} \textbf P_{\textbf q\nu }^{\kappa ,s }\delta(\omega-\omega_{\textbf q,\nu} ),
\end{split}
\label{projected-fw}
\end{equation}
for the atom type $s$ in the unit cell  (including In$_2$ or Se$_2$) where $ \kappa=\overline{xy} $ (labeled by in-plane), $ \overline z $ (labeled as out of plane), and 
\begin{equation}
\textbf P_{\textbf q, \nu}^{\overline{xy},s}= \sum_{\kappa=x,y} \textbf {e}_{\bf q,\nu  }^{*\,\kappa,s}\,\, \textbf {e}_{\bf q,\nu}^{\kappa,s}, \,\,\, \textbf P_{\textbf q, \nu}^{\overline{z},s}= \textbf{e}_{\bf q,\nu  }^{*\,z,s}\,\, \textbf{e}_{\bf q,\nu}^{z,s},
\end{equation}
where vector ${\bf e_{q\nu}}$ is the eigenvector of the dynamical matrix. The ${\alpha^2} \textbf{F}$ can also be projected into Cartesian coordinates by making use of the phonon displacements associated with various atom types in different directions,
\begin{equation}
\begin{split}
{ \alpha} ^2 {\textbf{F}} ^{\kappa, \kappa^\prime}_{s,s^\prime}(\omega) = \frac{1}{N_\textbf{k} N_{\textbf{q}} N(\varepsilon_F)} \sum_{m,n,\nu,\textbf{k}, \textbf{q}} g^{*\, \kappa,s}_{n\textbf{k},m\textbf{k+q},\nu} 
 \,\,  g^{\kappa^\prime, s^\prime}_{n\textbf{k},m\textbf{k+q},\nu}  \\ \times \,\, \delta(\varepsilon_{n\textbf{k}}-\varepsilon_{{F}})\delta(\varepsilon_{m\textbf{k+q}}-\varepsilon_{{F}}) \delta(\omega -\omega _{\textbf q, \nu}),
\end{split}
\label{projected-a2f}
\end{equation}	
where $\kappa,\, \kappa^\prime $ refer to the in-plane and out-of-plane deformations, respectively with $g^{\overline{xy}, s}_{n\textbf{k},m\textbf{k+q},\nu} =\sum_{\kappa=x,y} (\frac{\hbar }{2\omega_{\textbf{q}\nu}})^{1/2}\, d^{\kappa, s}_{n\textbf{k},m\textbf{k+q}}u^{\kappa, s}_{\textbf{q}\nu} $ and $g^{\overline{z}, s}_{n\textbf{k},m\textbf{k+q},\nu} = (\frac{\hbar }{2\omega_{\textbf{q}\nu}})^{1/2}\,d^{z, s}_{n\textbf{k},m\textbf{k+q}}u^{z, s}_{\textbf{q}\nu}$ and $u^{\kappa,s}_{\textbf{q}}= \frac{\textbf{e}^{\kappa,s}_{\textbf {q}}}{\sqrt{m_s}}$ is the displacement pattern \cite{Note1}, so that  ${ \alpha} ^2 {\textbf{F}} ^{\kappa, \kappa^\prime}_{s,s^\prime}$ satisfies the following relation 
\begin{equation}
\begin{split}	
	 { \alpha} ^2  {\textbf{F}}(\omega)  =  \sum_{k, k^\prime,s, s^\prime}  { \alpha} ^2  {\textbf{F}} ^{\kappa, \kappa^\prime}_{s,s^\prime}(\omega).
\end{split}
\label{projected-a2f-satisfied}
\end{equation}		
In particular, we define $\displaystyle \alpha^2  {\textbf{F}}_{\overline{z},\overline{xy}}(\omega)  =  2\sum_{s,s^\prime}\sum_{k^\prime=x,y}  \Re e[{ \alpha} ^2  {\textbf{F}} ^{z, \kappa^\prime}_{s,s^\prime}(\omega)]$, $\displaystyle   \alpha^2  {\textbf{F}}_{\overline{xy},\overline{xy}}(\omega)  =  \sum_{s,s^\prime}\sum_{k,k^\prime=x,y}  { \alpha} ^2  {\textbf{F}} ^{\kappa, \kappa^\prime}_{s,s^\prime}(\omega)$ and $\displaystyle\alpha^2  {\textbf{F}}_{\overline{z},\overline{z}}(\omega)  =  \sum_{s,s^\prime} { \alpha} ^2  {\textbf{F}} ^{z, z}_{s,s^\prime}(\omega)$.
Projected  $\lambda$ can be obtained by projected $\alpha ^2  {\textbf{F}} $ as follows:
\begin{equation}
\begin{split}	
 {\lambda} ^{\kappa, \kappa^\prime}_{s,s^\prime} = 2\int d\omega \frac{{ \alpha} ^2 {\textbf{F}} ^{\kappa, \kappa^\prime}_{s,s^\prime}(\omega)}{\omega}.
\end{split}
\label{projected-lambda}
\end{equation}

It would be worth mentioning the Fermi surface of monolayer InSe is anisotropic for some doping levels implying the importance of using the anisotropic Eliashberg theory. In this regard, the $\mu^*_c$ in anisotropic equations~\cite{PhysRevB.87.024505} was implemented as a cutoff independent quantity in EPW code. However,  to get better consistent 
 results comparable with that obtained by Eq.~\ref{eq:ADCORR}, we gain use of a cutoff dependent $\mu^*_N$  given by
\begin{equation}
\mu_{{N}}^{*}=\frac{\mu^*_{\mathrm{c}}}{1+\mu^*_{\mathrm{c}} \ln \left( \bar\omega_{2}/\omega_{N} \right)}
\label{muN}
\end{equation}
where $N$ represents the number of Matsubara frequencies at a defined temperature and $\omega_{N} \approx 8\, \bar\omega_2$ \cite{allen1975transition,Note2} is a good estimation. This approach provides better results compared with the one when $\mu^*_c$ is used~\cite{PhysRevB.87.024505}. The value of $T_c$ is obtained when the superconducting gap becomes smaller than $5\times 10^{-4}$ eV.
 
Furthermore, in metallic systems, the ion dynamic affects the electron dynamics and leads to the excited state owing to the proximity of phonon energies and electron excited states~\cite{calandra2010adiabatic}.
The experimental realization of such dynamics on the phonon energies is observable in the form of a Raman frequency shift at the zone center so-called non-adiabatic effects \cite{novko2019nonadiabatic, saitta2008giant, novko2020broken,giustino2017electron}.
To explore this, a time-dependent perturbation theory (TDPT) is necessary for a full {\it ab initio} treatment of non-adiabatic effects. Since a full TDPT is complicated enough in practical terms of complexity of the accurate calculations, we adopt the following procedure, by pursuing Ref. \cite{calandra2010adiabatic}, to properly include the non-adiabatic effects. As the first necessary step for a specific $ \textbf q$ vector, adiabatic self-consistent force constants, $C_{sr}(\textbf{q}$$, 0, T_{1})$, are calculated. Here $T_1$ is the electronic temperature applied in self-consistent calculations ($T_1$ is large enough to prevent a Kohn anomaly).
The non-adiabatic phonons can be naturally obtained by diagonalizing the phonon dynamical matrix related to non-adiabatic non-self-consistent force constants, $ \widetilde C(\textbf{q}$$, \omega, T_{0})$, at a physical temperature $T_0$ given by \cite{calandra2010adiabatic}:
\begin{equation}\label{eq10}
\widetilde C_{sr}(\textbf{q}, \omega, T_{0}) = \Pi_{sr}(\textbf{q}, \omega, T_{0})+ C_{sr}(\textbf{q}, 0, T_{1}).
\end{equation}
where $\Pi_{sr}$ comprises both the addition (subtraction) of non-adiabatic (adiabatic) effects at the specific temperature $T_0$ ($T_1$) used in the related Fermi-Dirac distribution function $f_{\textbf k m}$, respectively,
\begin{equation}\label{eq11}
\begin{split}
\Pi_{sr}(\textbf{q}, \omega, T_{0}) =
&\frac{2}{N_{\textbf k}(T_0)}\sum_{{\textbf k},m,n}^{N_{\textbf k}(T_0)} \frac{f_{{\textbf k} m}(T_0) - f_{\textbf{k+q} n}(T_0)}{\varepsilon_{\textbf k m}-\varepsilon_{\textbf{k+q} n}+\omega+i\eta} \times \\
&{\textbf d}^{s}_{mn}({\textbf k},{\textbf k}+{\bf q}) \, {\textbf d}^{r}_{nm}({\textbf k}+{\bf q},{\textbf k})\,\, \\ 
& - \frac{2}{N_{\textbf k}(T_1)}\sum_{\textbf k,m,n}^{N_{\textbf k}(T_1)} \frac{f_{\textbf k m}(T_1) - f_{\textbf{k+q} n}(T_1)}{\varepsilon_{\textbf k m}-\varepsilon_{\textbf{k+q} n}} \times \\
&{\textbf d}^{s}_{mn}({\textbf k},{\textbf k}+{\bf q}) \, {\textbf d}^{r}_{nm}({\textbf k}+{\bf q},{\textbf k})
\end{split}
\end{equation}
where $N_\textbf k(T_0)$ is the $\textbf k$-point grid at smearing $T_0$ and much larger than $N_\textbf k(T_1)$ and we consider $\eta$ as a positive real infinitesimal parameter. Furthermore, $ \textbf{d}^{s}_{mn}$ are deformation potential matrix elements which include the derivative of the Kohn-Sham self-consistent potential with respect to the Fourier transform of the phonon displacements \cite{calandra2010adiabatic}. Therefore, to obtain phonon energies within an adiabatic regime a coarse $24 \times 24 \times 1$ $\textbf k$-point mesh and $ T_{1} =$ 1580 K as a proper starting point are considered. While a dense enough $\textbf k$-point grid of $72 \times 72 \times 1$ is adopted for the calculation of non-adiabatic and adiabatic force constant matrices at more reduced temperatures  ($T_0$).

\begin{figure}[t]
\centering
\includegraphics[scale=1, trim={0cm 0cm 0cm 0cm }]{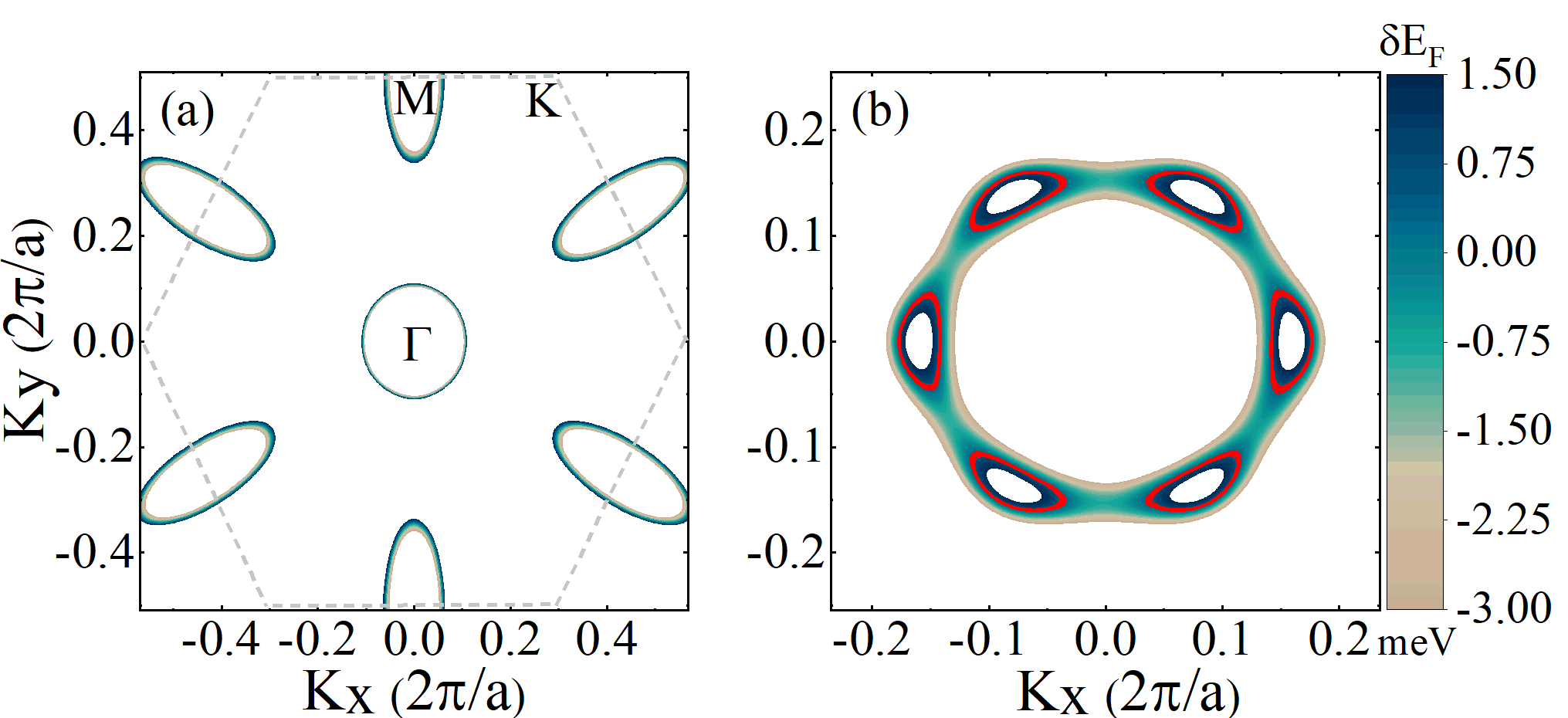}
\caption{(Color online) The Fermi surface contour of monolayer InSe based on the jellium model. (a) The Fermi surface for doping -0.1. (b) The Fermi surfaces corresponding to different shifts of the E$ _{\rm F}$ from the E$ _{\rm F}$ related to the doping level  +0.04 (represented by red lines). The color bar shows the shift of the Fermi energy. The gray dashed lines are applied to illustrate the first Brillouin zone boundaries. }

\label{Fsurface}
\end{figure}

\begin{figure}[]
\centering
\includegraphics[scale=0.420,trim={0cm 0cm 0cm 0cm}]{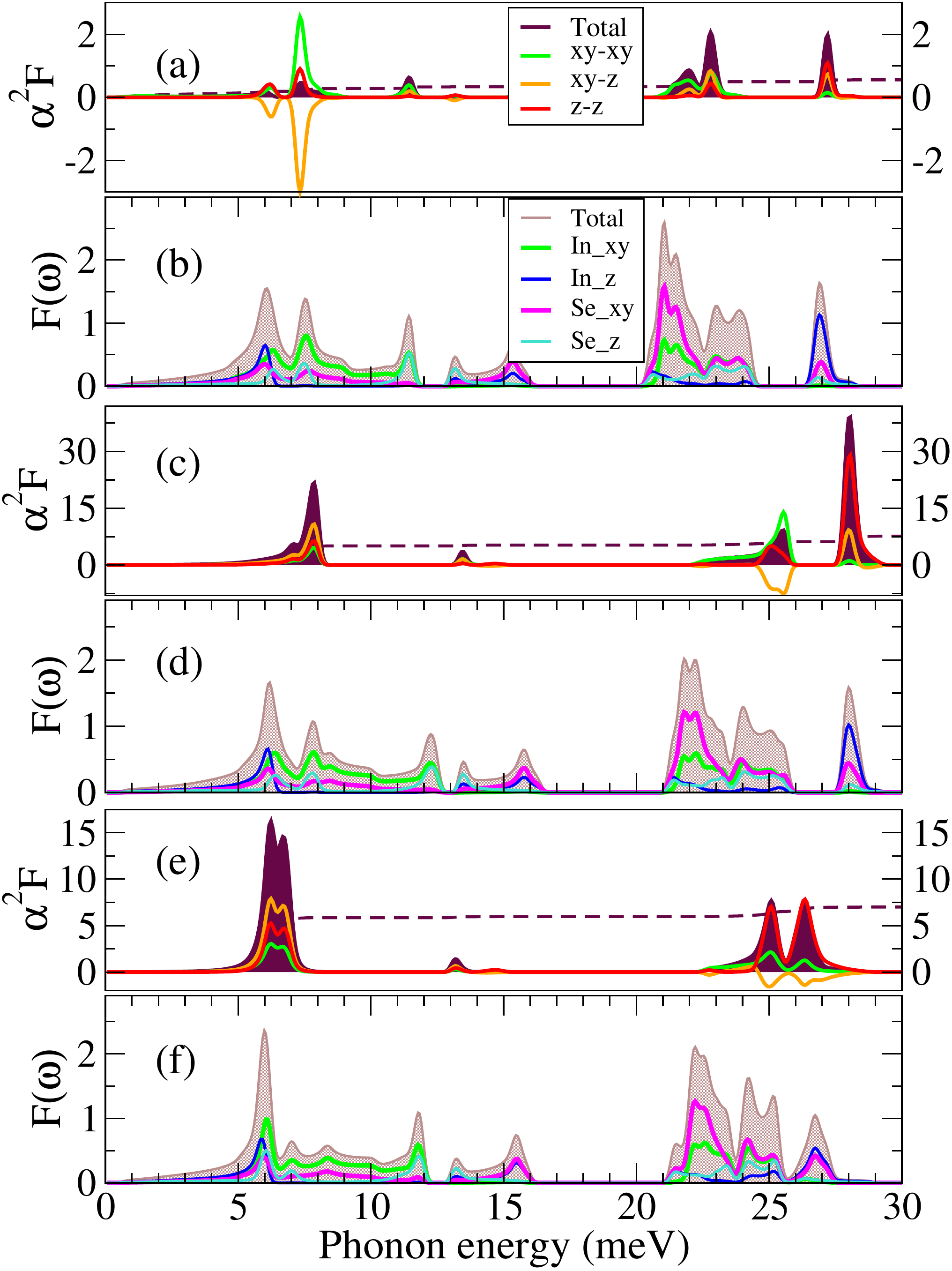}
\caption{(Color online) The $\alpha^{2}{\textbf{F}}$ and projected phonon DOS within the jellium model. (a), (c) and (e) Total $\alpha^{2}{\textbf{F}}$ and cumulative EPC, $\lambda(\omega)$,  for specific doping levels $-0.1 , +0.01$ and $+0.1$. The dashed lines are utilized for $\lambda(\omega)$. (b), (d) and (f) refer to the projected phonon DOS, $\textbf{F}(\omega)$, respectively, for In and Se atoms along the in-plane and out-of-plane directions for corresponding doping levels. All of the graphs have been plotted in $T_1 = 1580 $ K.}
\label{a2fphdos}
\end{figure}


\section{Numerical results and discussions}\label{sec:results}

Two distinct types of structural phases ($ \alpha$ and $ \beta $) have been properly reported for pristine monolayer InSe in Ref.~\cite{PhysRevB.89.205416} whose $ \alpha$ phase has mirror symmetry, while $\beta$ has inversion symmetry. Moreover, both of them are dynamically stable, but, the former possesses cohesive energy slightly lower than the latter. We efficiently perform our DFT calculations on $ \alpha $ phase by incorporating a hexagonal structure with $D_{3h}$ symmetry. The relaxed geometry calculations of pristine monolayer InSe show that the optimized hexagonal unit cell naturally has the lattice constant $a= 3.90$ {\AA} and two sublayers are separated by distance $d_{{In-In}}$ = 2.66 {\AA} and $d_{{Se-Se}}$= 5.15 {\AA}. These parameters are in good agreement with those results reported in \cite{PhysRevB.89.205416, hung2017two,chen2019phonon}.


\begin{table*}[t]
\caption{The superconducting properties of monolayer InSe including  the EPC constant, $\lambda_{{\text {tot}}}$, density of states at the
Fermi level, $ N(\varepsilon_F)$ in units of states/eV/spin/unitcell, logarithmically averaged phonon frequency, $\omega_{\text {log}}$, isotropic transition temperature to the superconducting phase, $T_c$, for the studied hole/electron concentrations. The charge density is in units of 10$^{13}$ cm$^{-2}$.  $T_c$s are calculated for three different amounts of $ \mu^*_c $ (0.1, 0.15 and 0.2).
\label{table1}}
\begin{tabular}{@{\extracolsep{\fill} }lccccccccccc}
\hline \hline
 e/f.u. & charge density & $  \bf{\bf\lambda_{\text {tot}}} $ \; & $ N(\varepsilon_F)$ \;  & $\omega_{\text {log}}$(K) \;  &  & \multicolumn{3}{c}{$T_c $(K)} \;  \\
   \cline{6-8}
 & & & & & $\mu^*_c=0.1$ & 0.15 & 0.20  \\
\hline
+0.01 \; &  0.758 \;& 7.62 \;  &4.26\; & 123\;   \; &65 \; &58\; &51   \\
+0.04 \; & 3.0 \;& 7.36 \;  &8.05\; & 106\;   \; & 55  \; &48\; &42  \\
+0.1 \; &  7.58 \;& 6.99 \;  &8.30\; & 90\;   \; &44 \; &38\; &34  \\
+0.2 \; & 15.1 \;& 3.07 \;  &3.02\; & 79\;   \; & 21  \; &17\; &15    \\
+0.3 \; & 22.6 \;& 1.44 \;  &1.50\; & 78\;  \; &9  \; &8\; &7   \\
 +0.4 \; & 30 \;& 0.85 \;  &0.76\; & 80\;   \; & 4   \; & 3 \; &2  \\
- 0.1 \; & 7.44 \; &0.55 \;  &0.82\; & 97\;    \; & 2 \; &1 \; &0   \\
- 0.2 \; &  14.6 \;& 0.50 \;  &0.63\; & 103\;   \; & 2  \;&1 \; & 0  \\
\hline \hline
\end{tabular}
\end{table*} 


\subsection{Investigation of superconductive properties of monolayer InSe}  
In this work, both the electron- and hole-doped cases are studied within the jellium model for monolayer InSe.  A compensate positive or negative background charge is included to guarantee the charge neutrality.\\
There are different experimental techniques like electrolytic gate
\cite{efetov2010controlling} to precisely control the rate of the electron and hole densities. Here, we consider electron doping levels $-0.1$ and $-0.2$ electron per formula unit (e/f.u.) precisely corresponding to the electron densities, $7.44\times10^{13} $ and $1.46\times10^{14} \,$ cm$^{-2}$ respectively. Similarly, $+0.01, +0.04$ (low doping regime), $+0.1, +0.2, +0.3$ and $+0.4$ e/f.u. (high doping regime) for hole-doped cases corresponding to $7.58\times10^{12}$, $3.0\times10^{13} $, $7.58\times10^{13} $, $1.51\times10^{14} $, $2.26\times10^{14} $ and $3.0\times10^{14} $ cm$^{-2}$ charge densities are considered. For the sake of simplicity, we promptly drop e/f.u. units corresponding to various doping levels, $+/-$ refers to the hole/electron doping, respectively.\\ 


The Fermi surfaces of the system are described in Fig.~\ref{Fsurface} for different doping levels. Figure~\ref{Fsurface}(a) displays the topology of the Fermi surface for $-0.1$ doping consisting precisely of two types of electronic pockets located at the $\Gamma$ and $M$ points. In the case of the deeper electronic doping level $-0.2$, the specific form of the Fermi surface is similar to the previous doping level. The Fermi surface of the $+0.04$ doping system consists of six separated pockets located around a point between the $\Gamma$ and K as shown in Fig.~\ref{Fsurface} marked by the red color.

In the hole-doped case [see Fig.~\ref{Fsurface}(b)] and upon more significantly decreasing the Fermi energy $E_{\rm F}$, a Lifshitz transition \cite{PhysRevB.89.205416} occurs. Therefore, the topology of the Fermi surface with six pockets, located between $\Gamma$ and K, changes to two coaxial pockets around the $\Gamma$ point. This fundamental change of the principal character of the Fermi surface results in a tangible variation of the superconductive properties of the hole-doped system which we adequately address in the following. Moreover, this specific concentration is obtained to be equal to 5.8$ \times 10^{13}\text{ cm}^{-2}$ or +0.076  e/f.u. which is in good agreement with that reported in Ref.\cite{PhysRevB.89.205416}.
To begin with, we carefully look at the Eliashberg function in terms of various doping levels.
Figures~\ref{a2fphdos}(a) and (b) depict the projected $\alpha^2 \textbf{F}(\omega)$ and phonon DOS for doping level $-0.1$. The projected Eliashberg function along the in-plane and out-of-plane deformations show a mighty peak at around 27 meV related to a scattering process which originates primarily from $\alpha^{2}{\textbf{F}}_{\overline{z},\overline{xy}} + \alpha^{2}{\textbf{F}}_{\overline{z},\overline{z}}$ resulting from the out-of-plane vibration of In atoms and in-plane vibration of Se atoms. This is equally consistent with the projected $ \textbf{F}(\omega)$ in Fig.~\ref{a2fphdos}(b), where there is a significant density of phonons with In$_z$ and Se$_{xy}$ deformations. Looking at more reduced energies there is a two-peak structure between $21-24$ meV, which comes from 
  ${\alpha^2\textbf{F}}_{\overline{z},\overline{xy}} + {\alpha^2\textbf{F}}_{\overline{xy},\overline{xy}}$. On the other hand, peaks at more reduced energies originate from ${\alpha^2\textbf{F}}_{\overline{xy},\overline{xy}}$.

The $\alpha^2\textbf{F}(\omega)$ and $ \textbf{F} (\omega)$
are shown in Figs.~\ref{a2fphdos}(c) and (d) for the low hole doping level $+0.01$. A peak around 28 meV comes principally from single optical phonon mode with out-of-plane In and in-plane Se vibrations. In this case, the deformation of ${\alpha^2\textbf{F}}_{\overline{z},\overline{z}}$ is considerably larger than ${\alpha^2\textbf{F}}_{\overline{z},\overline{xy}}$. Moreover, the lesser peak at around 26 meV has
a major ${\alpha^2\textbf{F}}_{\overline{xy},\overline{xy}}$ and a minor ${\alpha^2\textbf{F}}_{\overline{z},\overline{z}}$ character with a negative contribution from ${\alpha^2\textbf{F}}_{\overline{z},\overline{xy}}$, while the strong peak at around 8 meV has a major ${\alpha^2\textbf{F}}_{\overline{z},\overline{xy}}$ character with relatively similar contributions from the other two deformations.

As a notable example of high a hole-doped regime, the $\alpha^{2}{\textbf{F}}(\omega)$ and projected ${ \textbf{F}(\omega)}$ for $+0.1$ are shown in Figs.~\ref{a2fphdos}(e) and (f), respectively. Despite the low hole-doped and electron-doped cases, the prominent peak around 28 meV is absent.
In general, the spectrum of $+0.1$ hole-doped is slightly shrunk in comparison with the $+0.01$ one. Moreover, the gapped two-peak structure in the high-energy part of the $\alpha^2 \textbf{F}(\omega)$ for $+0.01$ is replaced with a gapless one at an energy of about 25-27 meV. The outstanding contribution of this high-energy part arises mainly from the ${\alpha^2\textbf{F}}_{\overline{z},\overline{z}}$ and ${\alpha^2\textbf{F}}_{\overline{xy},\overline{xy}}$ deformations, however, the ${\alpha^2\textbf{F}}_{\overline{z},\overline{xy}}$ has a completely negative contribution. The low-energy peak between 5-7 meV has almost an identical character to the low-energy peak of the $+0.01$ doping level, albeit with a lesser height.
Therefore, the peak of $\alpha^2{\bf F}$ is shifted to lower energies by passing through the Lifshitz transition point (increasing hole doping levels). In addition, there is a tangible suppression of the proportion of the spectral weight of high-energy phonons to low-energy phonons. Such a modulation of optical phonons affects their superconductive properties, which mainly manifests itself in the suppression of $\omega_{log}$ (see Table.~\ref{table1}). 

Looking at the cumulative $\lambda(\omega)$ in Figs.~\ref{a2fphdos}(a), (c) and (e), we can fairly state that in hole doping the acoustic branches carry out a more pronounced role in the formation of the $\lambda_{\text {tot}}$. Unlike the hole-doped cases, for electron doping, there is a more uniform distribution of each branch contributing in the formation of the $ \lambda_{\text {tot}}$ for the electron doping, as inferred from $ \lambda(\omega)$.\\
The tabulated amounts of $\lambda_{\text {tot}}$ with respect to various doped levels in Table.~\ref{table1} reveal that increasing the hole/electron doping levels leads to a descending/constant behavior of $\lambda_{\text {tot}}$, respectively.
\begin{figure}[!]
\centering
\includegraphics[scale=.3 ,trim = {0cm 0cm 0cm 0cm}]{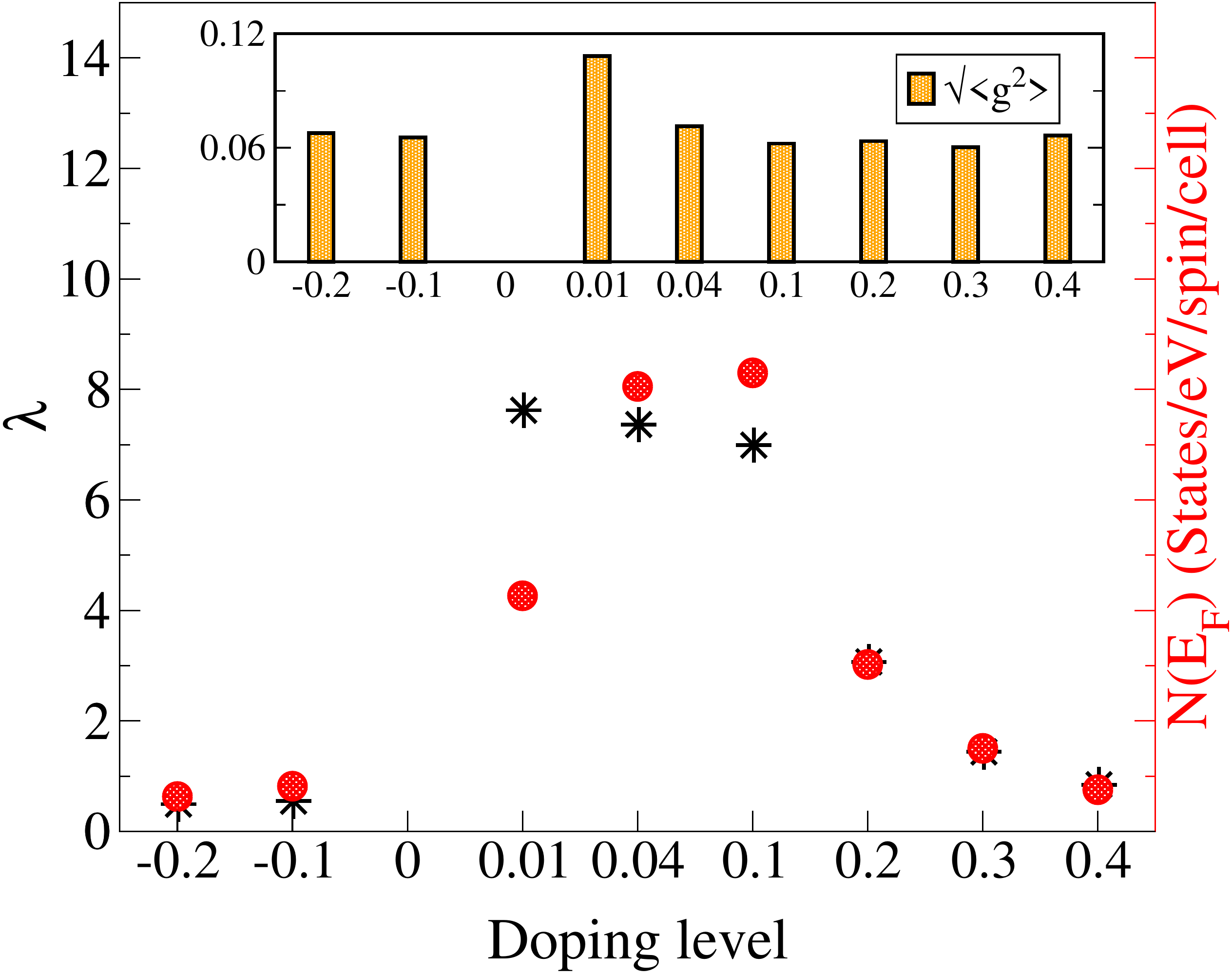}
\caption{ (Color online) The variation of the  EPC and $N(\varepsilon_{\rm F})$ with respect to different doping levels in monolayer InSe. Inset: The $ \sqrt{<g^2>}$ as a function of the doping level in units of eV. In the case of hole doping, an increment in the carrier density leads to a decreasing of both the $\lambda \text{\,\,and\, \,} N(\varepsilon_{\rm F})$ and nearly constant value for the $ \sqrt{<g^2>}$. While, in the case of electron doping, an increment in the carrier density leads to a constant behavior of $\lambda$, $  N(\varepsilon_{\rm F})\,\, \text{\,\,and\, \,} \sqrt{<g^2>}$. }
\label{lam-dos-g2}
\end{figure}
 To perceive the correlation between the 
 DOS at the Fermi energy $N(\varepsilon_{\rm F})$ and $\lambda$, we collect the results of Table.~\ref{table1} into Fig.~\ref{lam-dos-g2}, 
 where $\lambda$ and $ N(\varepsilon_{\rm F})$ are shown for different doping levels.
 Upon progressively increasing the hole density, while $\lambda$ decreases monotonously, the $N(\varepsilon_{\rm F})$ increases up to doping level $+0.1$ then decreases for a larger doping level. 
 
 One can seemly remark that $\lambda$ can take an effect from $N(\varepsilon_{\rm F})$ and the average of the electron-phonon matrix elements on the Fermi surface, and effectively could be represented as $ \lambda = 2N(\varepsilon_{F})\langle |g|^2\rangle /\omega_0$, where $\langle |g|^2\rangle$ is an average electron-phonon interaction. Therefore, if one considers the $\lambda/N(\varepsilon_{\rm F})$, it will be possible to recognize that the proportion is about unity for all doping levels but, +0.01 and 0.1,
  for doping level 0.01 an enhancement of the average electron-phonon interaction is expected.  To estimate the average electron-phonon interaction we use, $\displaystyle \langle|g|^2\rangle=\frac{1}{N(\varepsilon_{\rm F})}\int \alpha^2 \textbf{F}(\omega) d\omega$, and the results of the $\langle |g|^2 \rangle$ are presented in the inset of Fig.~\ref{lam-dos-g2}. As seen, the average electron-phonon interaction is enhanced for +0.01, compared to the other hole and electron doping levels. Thus, in general, a larger DOS results in a larger $\lambda$ with a linear dependency, with the only exception being the 0.01 doping level, where $\langle|g|^2\rangle$ is enhanced in comparison with the other doping levels where it is almost constant. \\

\begin{figure}[!]
\centering
\includegraphics[scale=.33,trim = {0cm 0cm 0cm 0cm}]{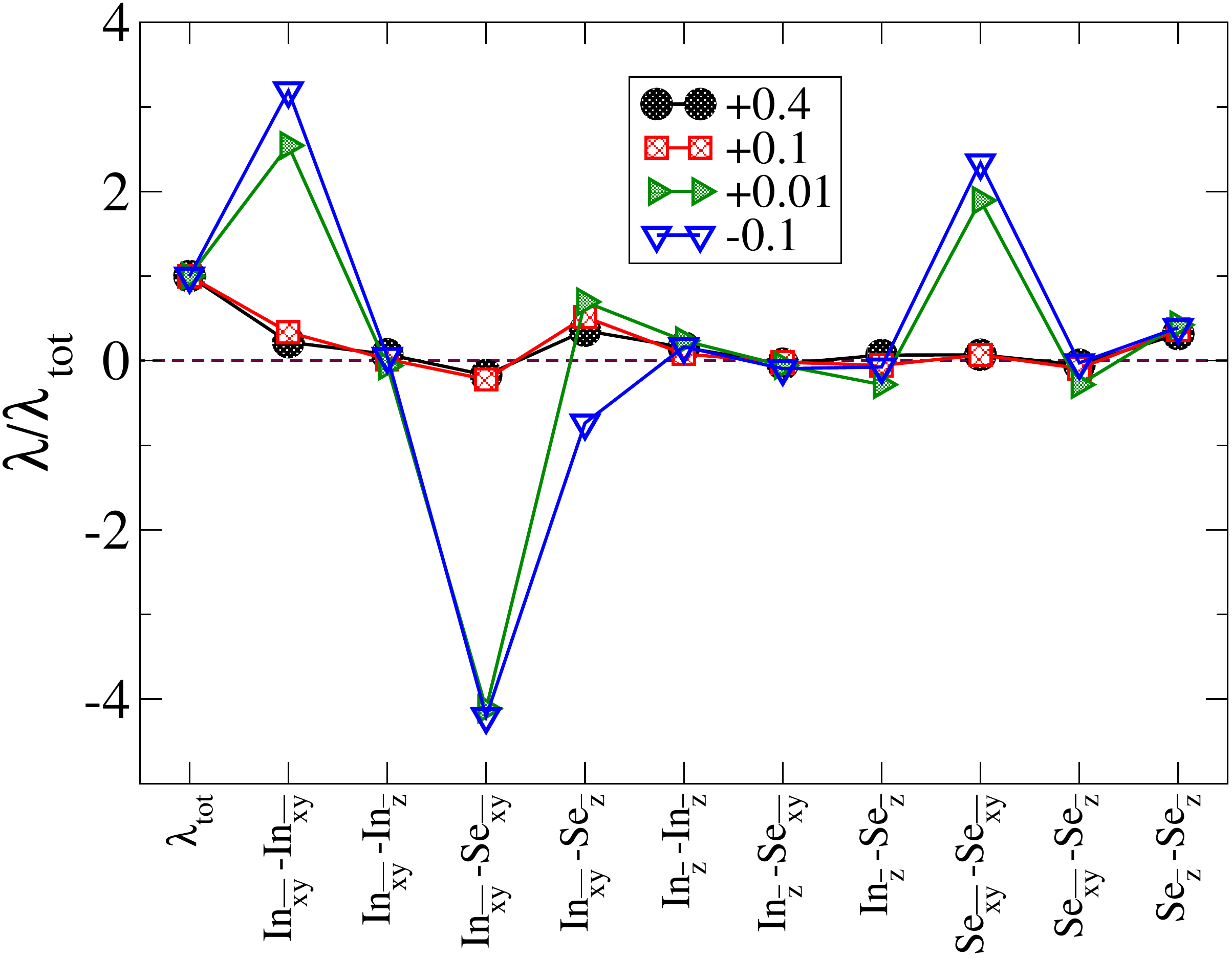}
\caption{ (Color online)  The projected $\lambda $  associated with different displacements of atoms along with the in-plane (xy)  and out-of-plane (z) directions for doping $+0.4, +0.1, +0.01$, and $-0.1$. The $\lambda$ parameter is rescaled to $\lambda_{{\text {tot}}}$. The dashed maroon line set to zero indicates the positive or negative role of the polarization. The splines connecting the points are used to guide the eyes.}
\label{lambda-doping-proj-rescale-Inin-Inin}
\end{figure}

Furthermore, Eq.~(\ref{projected-lambda}) is used to carefully consider the contribution of the projected $\lambda$ for different atom types and the out-of/in-plane directions in the $ \lambda_{\text {tot}}$.
Figure~\ref{lambda-doping-proj-rescale-Inin-Inin}  shows the projected $\lambda$ while those are rescaled to the
$\lambda_{\text {tot}}$ for four doping levels $-0.1, +0.01, +0.1$ and $+0.4$. The desired results show, for the electron-doped case, the highest contribution to the $\lambda_{\text {tot}}$ is attributed to the in-plane displacements. For $-0.1$ the corresponding in-plane/out-of-plane contributions are $ {\lambda_{\overline{xy},\overline{xy}}}=0.73 > \lambda_{\overline{z},\overline{z}} = 0.27>\lambda_{\overline{z},\overline{xy}}=-0.45$.
 
For the hole-doped levels beyond +0.1, on the other hand, the largest contribution arises from the out-of-plane deformations and mixed in-plane In and out-of-plane Se  deformations. For doping level +0.1 the projected $\lambda$s read, $ {\lambda_{\overline{z},\overline{z}}}=2.73 > \lambda_{\overline{z},\overline{xy}} = 2.7>\lambda_{\overline{xy},\overline{xy}}=1.55$.

In the case of +0.01 doping, the system is somewhere between a greater hole doping and the electron-doped
cases. While its in-plane contributions share the same behavior as of the electron-doped one; its out-of-plane and mixed in-plane/out-of-plane contributions behave properly similar to the high doping levels, $ {\lambda_{\overline{z},\overline{z}}}=2.88 > \lambda_{\overline{xy},\overline{xy}} = 2.48>\lambda_{\overline{z},\overline{xy}}=2.26$.
To be specific, the valuable contribution which comes from (In$_{\overline{xy}}\,$-Se$_{\overline{z}}$) deformation has a negative impact for the electron-doped system, while it has a positive contribution for low and high hole-doped cases.
This key difference originates from the distinction between the generic forms of the topology of the Fermi surface such that this type of polarization is beneficial for hole-doped cases and it is a disadvantage for the electron-doped ones.\\


In addition, Table~\ref{table1} shows the critical transition temperature to the superconducting phase with the aforementioned doped conditions calculated through Eq.~(\ref{eq:ADCORR}) by considering three values for $\mu^*_c = 0.1, \,\,0.15\text{ and }0.2 $. In the case of hole doping, the highest value of $T_c= 65$ K is obtained for $\mu^*_c = 0.1$. Our results reveal  that $T_c$ can be shrunk about 20 percent when  $ \mu^*_c = 0.2$ was applied. 
Obviously, while the amount of $\lambda$ is almost the same for the first three hole-doped cases, the $T_c$ for $+0.01$ is larger than that of $+0.1$ (by considering $ \mu^*_c = 0.1$); stemming from a larger value of $\omega_{\text {log}}$. The larger value of the $\omega_{\text {log}}$ corresponding to the former originates from the fact that the phonon dispersion for $+0.01$ doped is typically harder than $+0.1$. Moreover, the proportion of the high energy peak to the low-energy peak of $\alpha^2 \textbf{F}$ for the case of +0.01 is appreciably larger than that of $+0.1$ (see Figs.~\ref{a2fphdos}(c) and (e)). Thus, $\omega_{\text {log}}$ is enhanced for $+0.01$ in comparison with $+0.1$.\\

\begin{table}[t]
\caption{The anisotropic superconducting transition temperature of the monolayer InSe. The transition temperatures to the CDW region, $T_{CDW}$, in both adiabatic (A) and non-adiabatic (NA) regimes were obtained by using the fitting curve.  Three various amounts of $ \mu^*_c $ (0.1, 0.15 and 0.2) are used to calculate $T_c$.
\label{table2}}
\begin{tabular}{@{\extracolsep{\fill} }lcccccccccc}
\hline \hline
 e/f.u. & \multicolumn{3}{c}{$T_c $(K)} \;   & $T_{CDW}^{A} $(K) \;& $T_{CDW}^{NA} $(K) \;\\
  \cline{2-4}
 & $\mu^*_c=0.1$ & 0.15 & 0.20  \\
\hline
+0.01 \; &73 \; &64\; &54  \;\; &  122\; & $<$ 2    \\
+0.04 \; & 75  \; &68\; &62  \;\; &  145\; &  $<$ 2 \\
+0.1 \; &55 \; &50\; &43  \; \; &  416\; & 120  \\
+0.2 \; & 20  \; &17\; &15  \; \; &  539\; &191 \\
+0.3 \; &9  \; &8\; &7  \;\; &  476\; &246    \\
 +0.4 \; & 4   \; & 3 \; &2  \;\; &  $<$ 2\; & $<$ 2  \\
- 0.1 \; & 2 \; &0 \; &0  \;\; &  $<$ 2\; & $<$ 2  \\
- 0.2 \; & 2  \;&1 \; & 0 \;\; &   $<$ 2\; & $<$ 2 \\
\hline \hline
\end{tabular}
\end{table} 

Notice that the highest tabulated temperature is comparable with $T_c = 88$ K for blue phosphorene studied in Ref.~\cite{esfahani2017superconducting}. Moreover, it is much larger than the reported $T_c$ for Li-decorated monolayer graphene and antimonene with $T_c\approx 6 $~\cite{ludbrook2015evidence} and $ 4 \,$ K \cite{lugovskoi2018electron}, respectively. However, the high value of $\lambda$ needs a careful examination and further insights into the formation of the CDW phase at low temperatures for the hole-doped system which we adequately address in the following section. 

To have a better estimate of $T_c$, we utilize a self-consistent solution of the anisotropic the Migdal-Eliashberg theory. The results are reported in Table.~\ref{table2}. Obviously, anisotropic effects alter $T_c$ at the first three hole-doped cases, where the Fermi surface has a 
 more pronounced anisotropic character (see Fig.~\ref{Fsurface}(b)), while a slight variation of $T_c$  is observed for other hole- and electron-doped cases when $T_c$s are extracted from Allen-Dynes formula ( Eq.~\ref{eq:ADCORR}) and self-consistent anisotropic Eliashberg equations.
  These results indicate that below the Lifshitz transition point, in comparison with the Allen-Dynes estimate, the $T_c$ is more pronounced in comparison with the cases above the Lifshitz transition point as well as the electron doped levels. For the +0.04 doping level, such an anisotropy can enhance $T_c$ from a range 42-55 K corresponding to the Allen-Dynes estimate to 62-75 K for different applied $\mu^*_c$.


\subsection{CDW formation in adiabatic and non-adiabatic approximations}\label{sec:CDW}

\begin{figure}[!]
\centering
\includegraphics[scale=.35,trim = {0cm 0cm 0cm -1cm}]{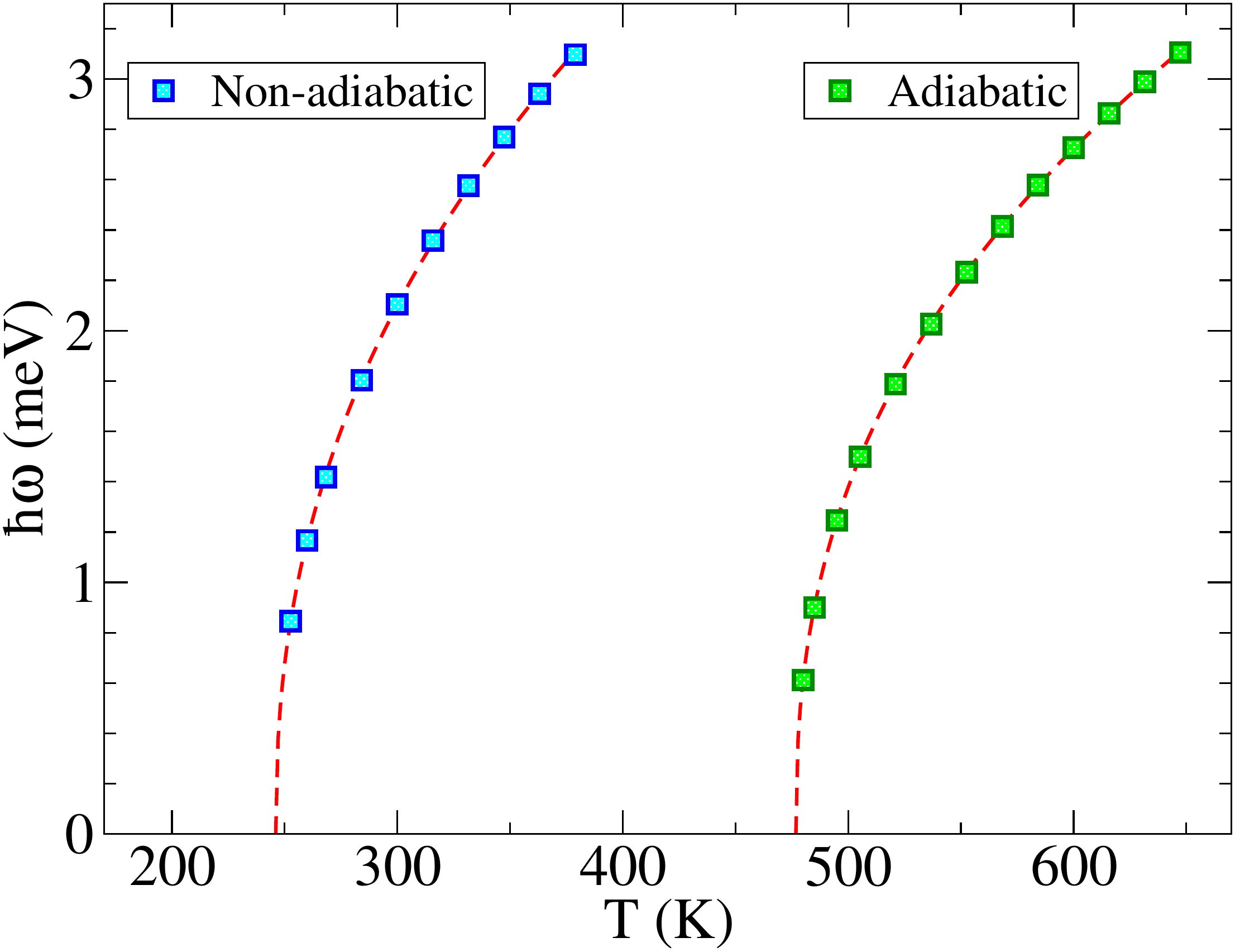}
\caption{ (Color online) Estimation of $T_{CDW}$ for case +0.3 in both adiabatic and nonadiabatic regimes  with respect to  different electronic temperatures. The red dashed lines illustrate the mean-field fitting according to $\omega = a_0\,(T-T_{CDW})^\delta$. Here, $a_0$  and $\delta$ yield values about 0.37 and 0.41, respectively.}
\label{fitting_tc}
\end{figure}

More reduction in the electronic temperature to achieve $T_{CDW}$ is alongside the giant amplitude of the Kohn anomaly. To acquire an estimate of $T_{CDW}$, we extract the frequency of the most softened mode on the whole $q$-mesh, for different temperatures, then we fit the extracted frequencies to 
 $\omega = a_0 (T-T_{CDW})^\delta$\cite{PhysRevB.92.245131}. In our calculations, $a_0$ is a constant close to 0.4 and $\delta$ yields values in the range 0.40-0.43 for all hole doping levels which are partly close to the value $\delta=0.5$ extracted from the mean-field approximation \cite{PhysRevB.92.245131,gruner2018density}.  Figure~\ref{fitting_tc} shows the variation of the phonon frequencies as a function of electronic temperature and related fitting curves (red dashed lines) for case +0.3 in both adiabatic and non-adiabatic regimes. The results indicate that the transition to the CDW region occurs in $T_{CDW}^A = 476 $ K and $ T_{CDW}^{NA}= 246 $ K. The values of $T_{CDW}$ corresponding to other doping levels, for both adiabatic and non-adiabatic regimes are reported in Table~\ref{table2}.   

\begin{figure}[t]
\centering
\includegraphics[scale=.33,trim={0cm 0cm 0cm 0cm} ]{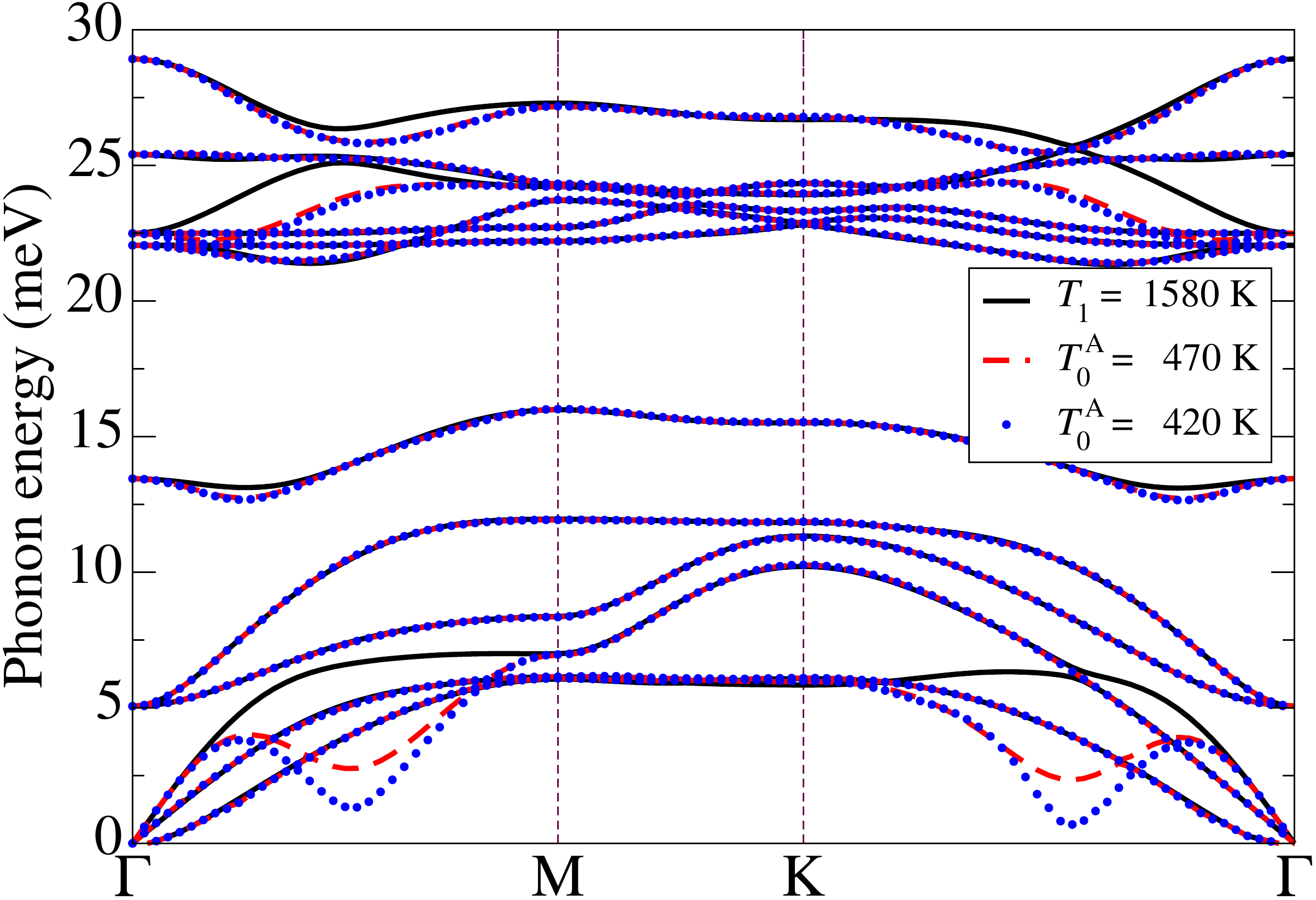}
\caption{ (Color online) The phonon dispersion as a function of electronic temperatures for a doping +0.1  in  the adiabatic (A)  regime. The lower the electronic smearing, the appearance of a greater amplitude of the Kohn anomaly which finally leads to CDW instability at a temperature lower than $T_0 =$ 416 K in the adiabatic regime while these instabilities can be faded out at upper temperatures such as $ T_0 =$ 420 and 470 K.   The black solid lines were carried out with typical electronic broadening, $T_1 =$ 1580 K, which is large enough to wipe out the Kohn anomaly in linear response self-consistent force constants.}
\label{CDW-temp}
\end{figure}

Figure~\ref{CDW-temp} depicts the amplitude of the Kohn anomaly as a function of the electronic Fermi-Dirac smearing for doping level +0.1. Typically decreasing the temperature leads to a more softening of the phonon energies and finally, the system suffers from a CDW instability at a smearing slightly lower than 416 K. For exploring the considerable variations of the phonon softening as a function of the Fermi-Dirac smearing, three upper temperatures, 420, 470 and 1580, in the adiabatic/static regime are depicted. The typical smearing 1580 K, as a starting point in the adiabatic regime, is large enough to wipe out the Kohn anomaly in the linear response calculations.
In addition, this figure shows there are two $\textbf q_{CDW}$ which give rise to two different chiralities. One includes a $6\times6$ commensurate supercell corresponding to the dip in the middle of the $\Gamma$-$K$ direction. The secondary point of the CDW instability is related to an incommensurate distortion precisely corresponding to another dip along the $\Gamma$-$M$ path.
Our numerical calculations reveal that the dip in the middle of the $\Gamma$-K direction has a lower $\omega$ and we, therefore, refer to this point as $\textbf q_{CDW}$ in the reminder. Notice that for the other higher doped levels, i.e. $+0.2$ and $+0.3$, the CDW forms at the same $\textbf {q}$ for the +0.1 doping level. On the other hand, in the adiabatic regime, low hole doping levels +0.01 and +0.04 show instability in a $\textbf {q}$ marginally different from the high doped regime. However, it does not show any instability of the system even at extremely low temperatures by including non-adiabatic effects as illustrated in Table~\ref{table2}. Besides, in the comparison between low doped and high doped regimes in terms of phonon softening at $\textbf q_{CDW}$, we therefore report our results at $\textbf q_{CDW}$ for doping levels $+0.01$ and $+0.04$ as well. 

\begin{figure}[!]
\centering
\includegraphics[scale=.30,trim={0cm 0cm 0cm 0cm} ]{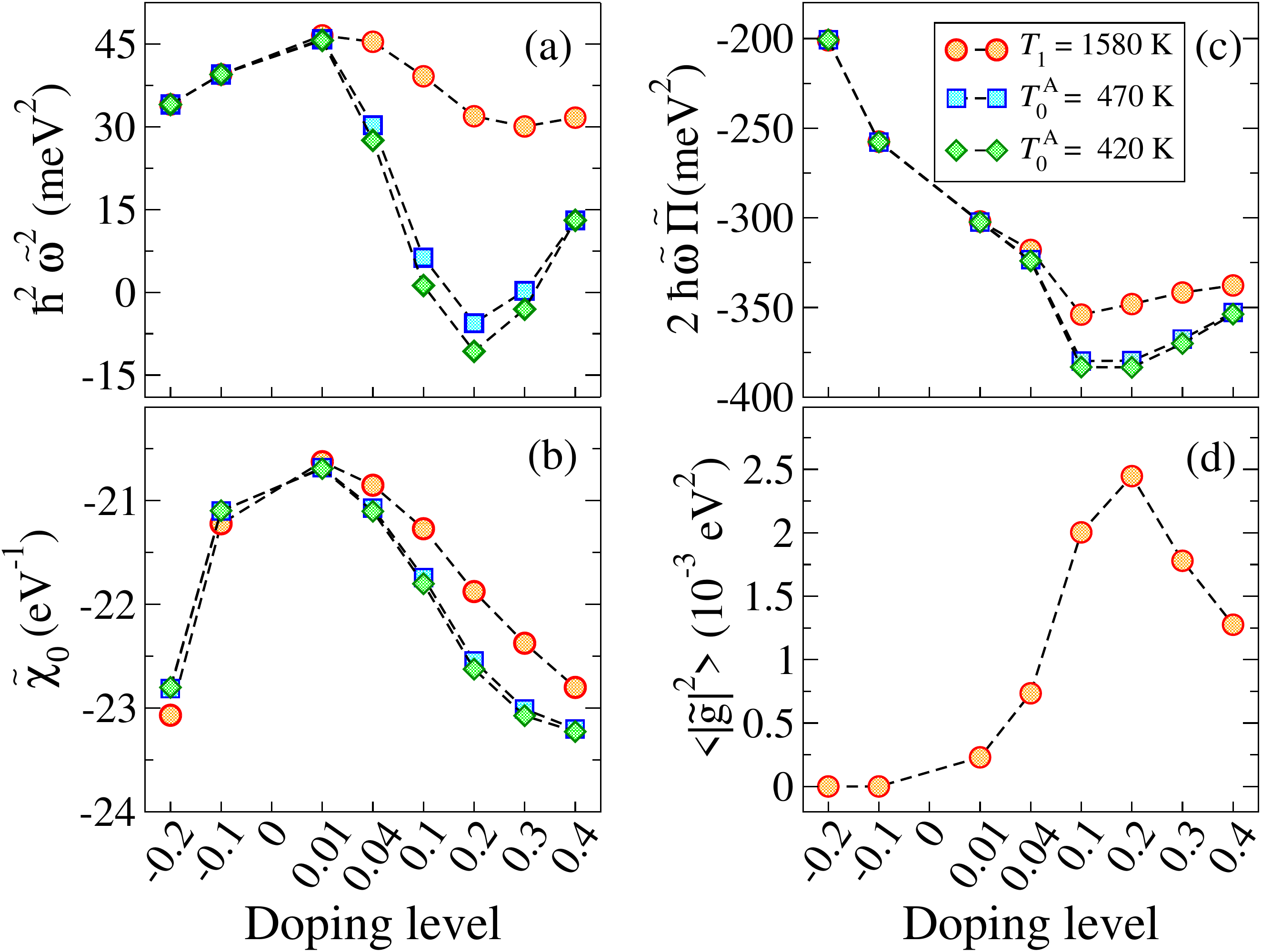}
\caption{ (Color online) The effective factors to determine the CDW instability as a function of temperature for all doped levels in the  adiabatic regime. (a) The dressed phonon energies squared, (b) electronic bare susceptibility, (c) real part of the phonon self-energy multiplied by $2\omega$,  and (d) the magnitude of $ \langle |\textbf{\~g}|^2\rangle$ related to the softening branch of phonon dispersion. The splines connecting the points are guides to the eyes and the tilde symbol refers to related calculations at $\textbf{q}_{CDW}$.}
\label{CDW}
\end{figure}

Figure~\ref{CDW} shows different quantities associated with the CDW formation for various doping levels and temperatures. In particular, the average amounts of the electron-phonon interaction $\langle \textbf{g}^2\rangle_{\textbf {q}\nu}= \frac{\gamma_{\textbf {q}\nu}}{2\pi\omega_{\textbf {q}\nu}\xi_{\bf q}}$, where the nesting function is precisely defined as $ \xi_{\bf{q}}= {N}^{-1}_{{\bf k}} \sum_{mn,{\bf k}} \delta(\varepsilon_{n{\bf k}}-\varepsilon_F)\delta(\varepsilon_{m{\bf{k}}+{\bf{q}}}-\varepsilon_F)$, is properly used.
The tilde symbol in Fig.~\ref{CDW} refers to the related calculations at the $\textbf{ q}_{CDW}$. Moreover, the depicted quantities are associated with the softened branch at $\textbf{q}_{CDW}$, therefore, the branch index $\nu$ is dropped.

The effects of phonon energy renormalization as a function of temperature within the adiabatic/static regime are shown in Fig.~\ref{CDW}(a). These results reveal the tendency of the system to the CDW region for the three $+0.1,\; +0.2$, and $+0.3$ doping levels. On the contrary, the electron-doped and low hole doping levels, below the Lifshitz transition point, almost retain their constant behavior as a function of various temperatures.
Figure~\ref{CDW}(b) shows the bare susceptibility as a function of doped levels at the $\textbf{q}_{CDW}$ for the aforementioned temperatures.
Notice that the $ \langle{\tilde{\textbf g}}^2\rangle$ is the largest for doping level $+0.2$ (see Fig.~\ref{CDW}(d)), in addition, the largest change of the $\chi_0$ basically belongs to the doping level $+0.2$. This leads to a further decline of $2\tilde{\omega}\tilde{\Pi}$ (from a temperature of 1580 K) for doping level $+0.2$ as shown in Fig.~\ref{CDW}(c). Moreover, such a larger variation in the $\chi_0$ for doping levels $+0.1,\; +0.2$, and $+0.3$ leads to a giant Kohn anomaly and finally the appearance of the instability in monolayer InSe for smearing lower than 416, 539 and 476 K, respectively. A comparison for doping $+0.4$ implicitly expresses that though there is a reduction of the self-energy correction, having less temperature dependence on $\chi_0$ together with a smaller
average of $ \langle{\tilde{\textbf g}}^2\rangle$ (Fig.~\ref{CDW}(d)) on the Fermi surface, results in less effective Kohn anomaly and therefore, the CDW is suppressed at $ \textbf q_{CDW}$ for doping level $+0.4$.

Further analyses associated with the polarization of the softened mode at $\textbf q_{CDW}$
adequately explain the instability at this point mainly involves the in-plane displacements of the In atoms and the out-of-plane displacements of the Se atoms at the same time.

The notable absence of the Kohn anomaly for an electron doping is owing to the lack of a reduction of the $\chi_0$ with respect to the different temperatures alongside an extremely small $ \langle{\tilde{\textbf g}}^2\rangle$ (Fig.~\ref{CDW}(d)).
In two low hole doping cases, $ \langle{\tilde{\textbf g}}^2\rangle$ is smaller than that obtained for other hole-doped levels. For doping level $+0.01$ a specific combination of a small $ \langle{\tilde{\textbf g}}^2\rangle$ and the lack of typically decreasing of $\chi_0$ as a function of temperature results in the absence of the Kohn anomaly at ${\bf q}_{CDW}$. In doping level $+0.04$, although there is a depletion in $\chi_0$ upon temperature reduction, due to a slight value of $ \langle{\tilde{\textbf g}}^2\rangle$, it sufficiently shows a smaller softening.
 \begin{figure}[t]
\centering
\includegraphics[scale=.35,trim={0cm 0cm 0cm 0cm} ]{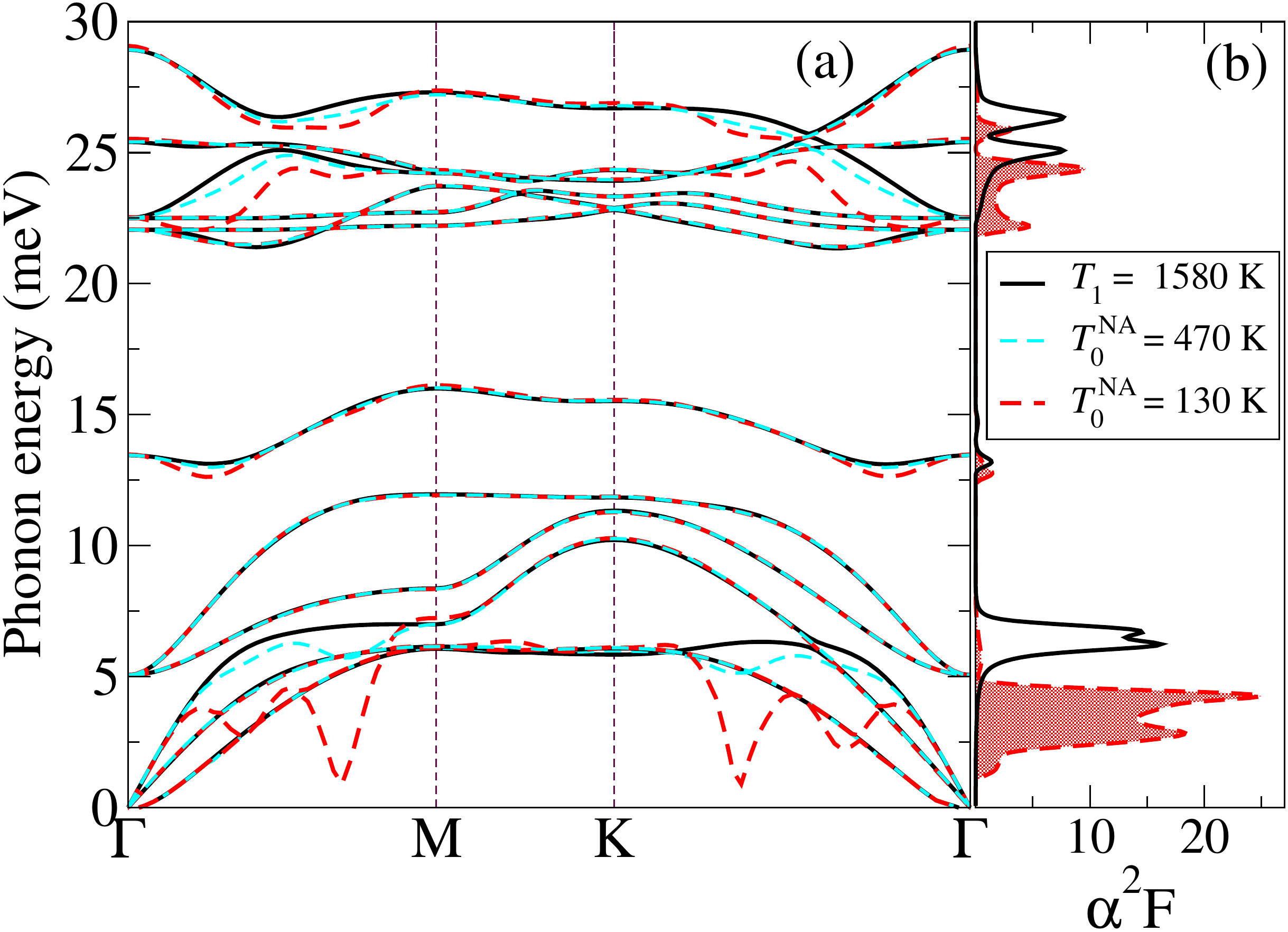}
\caption{ (Color online) The phonon dispersion corresponding to adiabatic high ($T_1$ = 1580 K) and  non-adiabatic low ($T_0$ = 130 and 470 K) electronic smearing for  doping +0.1. The related Eliashberg spectral functions are depicted at the right side of the plot for two temperatures $T_0$ = 130 and $T_1$ = 1580 K. Applying the non-adiabatic effects well expresses the suppression of the CDW phase  at a low temperature of 130 K. }
\label{CDW-temp2}
\end{figure}   
Therefore, considering the adiabatic regime, competition and coexistence between $T_c$ and $T_{CDW}$, reveals that $T_{CDW}$ is exceedingly greater than $ T_c$ and consequently, the CDW instability prevents access to the high-temperature superconductivity in the first five hole-doped cases +0.01, +0.04, +0.1, +0.2 and +0.3.   
On the other hand, in the intra-sheet scattering process, when $ |\varepsilon_{\textbf{k+q} }- \varepsilon_{\textbf k }| \approx\omega$, the substantional difference of the non-adiabatic and adiabatic frequencies is $\Delta \omega$ which approximately specifies $\Delta \omega \simeq N(\varepsilon_{\rm F}) \langle{\tilde{\textbf g}}^2\rangle$ at the Fermi surface \cite{novko2019nonadiabatic, saitta2008giant, lazzeri2006nonadiabatic}. Hence, this proper discrepancy is remarkable for the doping cases $+0.01, +0.04, +0.1, +0.2$ and $+0.3$ encompassing large amounts for both the $N(\varepsilon_{\rm F}$) and $ \langle{\tilde{\textbf g}}^2\rangle$; essentially restating the considerable importance of the non-adiabatic effects for these hole-doped cases.

 Figure~\ref{CDW-temp2}  shows non-adiabatic effects on phonon modes in the case of $+0.1$ doping for two low temperatures ($T_0$ = 130 and 470 K) together with a high enough temperature ($T_1$ = 1580 K). In order to perceive the effect of the phonon softening on $T_c$,  $T_0$ = 130 K is chosen such that it is slightly larger than $ T_{CDW}^{NA}$ = 120 K. Employing non-adiabatic phonons at $T_0$ = 130 K for the calculation of $T_c$ results in a slight enhancement of $T_c$ = 57 K within anisotropic Eliashberg theory, which still is much smaller than $ T_{CDW}^{NA}$ = 120 K. This lack of enhancement of $T_c$ could be understood based on Allen-Dynes estimation of $T_c$, as  softening related to phonon modes is accompany with shift of $\alpha^{2}  \textbf{F}$ to the lower frequencies, $T_{0} $ = 130 K, in particular in acoustic branches (see Fig.\ref{CDW-temp2}(b)). 
 This softening results in both remarkable enhancement of $\lambda $ and suppression of $\omega_{\text{log}}$   at the same time which finally leads to a little enhancement of $T_c $. Notice that the amplitude of the Kohn anomaly decreases in the presence of non-adiabatic effects as one may compare the phonon dispersion corresponding to electronic broadening at $ 470$ K in Figs.~\ref{CDW-temp} and \ref{CDW-temp2}. 

 \begin{figure}[t]
\includegraphics[scale=.3450,trim={0cm 0cm 0cm 0cm} ]{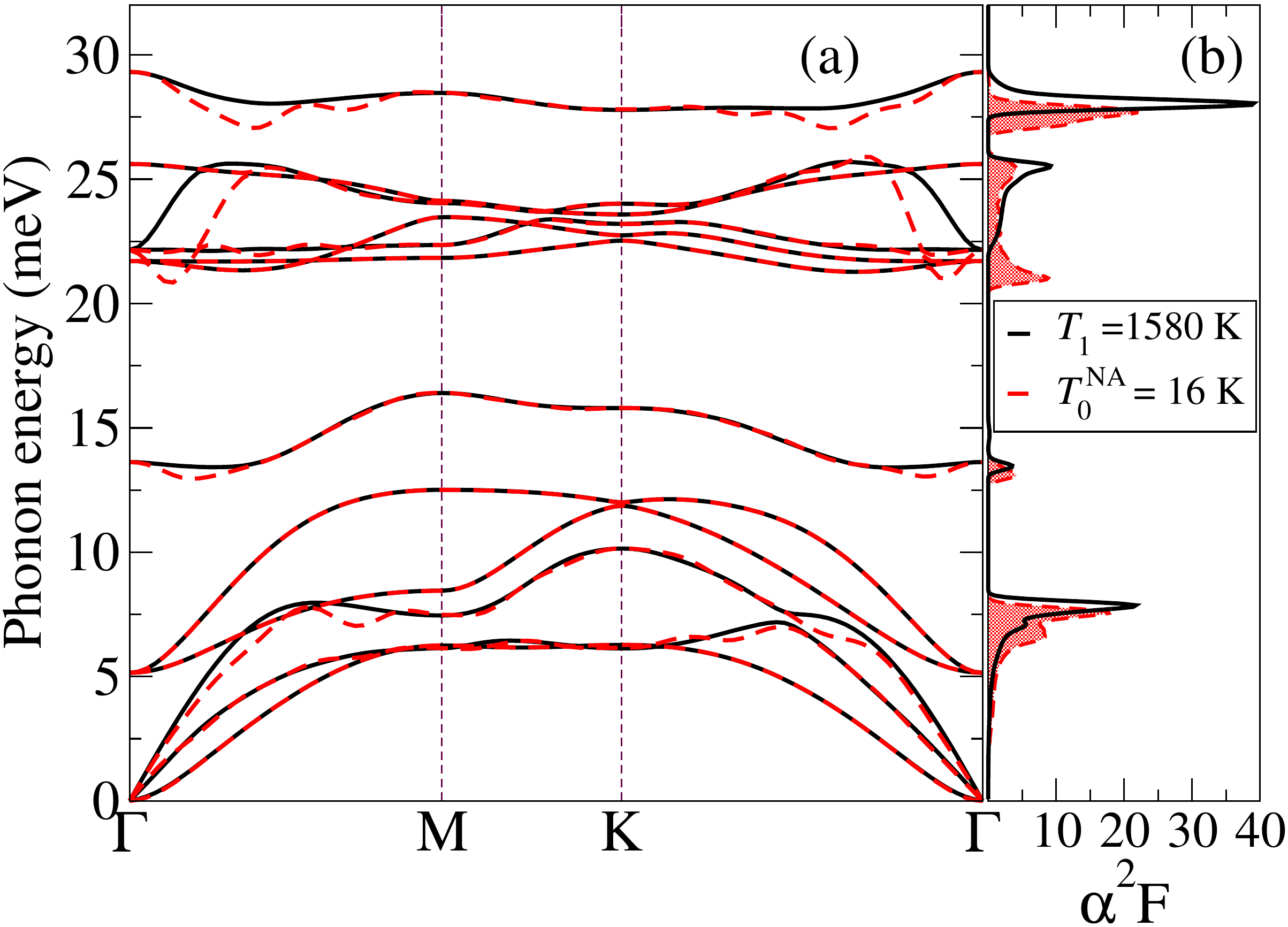}
\caption{ (Color online) The phonon dispersion corresponding to adiabatic high ($T_1$ = 1580 K) and  non-adiabatic low ($T_0$ = 16 K) electronic temperatures for  doping +0.01. The corresponding Eliashberg function is depicted on the right side of the plot.  Also, fading out of the CDW phase via non-adiabatic phonons at low temperatures is visible.}
\label{phonon-tcdw}
\end{figure}

In addition, the same calculations are repeated for doping $+0.2$ and $+0.3$. Applying the non-adiabatic effects on the phonon modes in two cases $+0.2$ and $+0.3$  at temperatures slightly above their $T_{CDW}^{NA}$  reveal a negligible enhancement of the superconducting transition temperature, $T_c$ = 26, 15 K respectively. This slight enhancement of $T_c$
  is simultaneous with a considerable suppression of $\omega_{\text{log}} $ = 36, 36 K respectively.

Consequently, non-adiabatic effects shift only the CDW region to lower temperatures 120, 191 and 246 K for elevated doping levels  $+0.1, +0.2$ and $+0.3$, respectively, and are not capable of suppressing the formation of the CDW instability in these three cases. Therefore, it appears that the superconducting transition for the three mentioned hole doping levels is unlikely to be accessible as a CDW phase forms before a superconductive phase. On the other hand, Table~\ref{table2} shows no dynamic instability at the remaining doped levels in the presence of non-adiabatic effects.

In Fig.~\ref{phonon-tcdw} the high temperature phonon dispersion, $T_{1} $ =  1580 K, and non-adiabatic low temperature one with $ T_0^{NA} = 16$ K are plotted along with their corresponding $\alpha^2\textbf{F}$ for hole doping level +0.01. The system is stable even for temperatures considerably smaller than its $T_c\, \approx$  54-73 K (see  Table~\ref{table2}). Notice that the $\alpha^2\textbf{F}$ calculated based on non-adiabatic phonons gives marginally different $T_c$ as small as 2 K, owing to the slight softening at certain $\textbf q$ points. Accordingly, the low hole-doped monolayer InSe likely shows a superconductive phase with maximum $T_c \sim 75$ K. The same analysis holds for hole doping level +0.04, where $T^{NA}_{CDW}$ is far below  its $T_c$ as it is shown in Tables \ref{table1} and Table \ref{table2}.

Note that the convergence of Eq. (\ref{eq11}) for $\eta$ is carefully checked to adequately explain this equation becomes practically $\eta$-independent when $\eta$ was changed in the range of 0.0015-0.015 Ry. In addition, the desired results reported in Table~\ref{table2} show that in the presence of the non-adiabatic effects, monolayer InSe is dynamically stable for all aforesaid doped levels at room temperature because all $T_{CDW}^{NA} $s are lower than room temperature.


\section{Conclusion}\label{sec:conclusion}

In summary, based on the first-principles DFT and DFPT methods, the superconducting properties of pristine monolayer InSe employing the Migdal-Eliashberg theory are explored. We have also calculated the renormalized phonon dispersion owing to the electron-phonon coupling in both the adiabatic and non-adiabatic regimes for various temperatures and doping levels. We have further investigated the competition between CDW formation and the superconductive phase for various hole and electron doping levels.

We have adequately discussed the most important phonon wave vectors leading to the remarkable electron-phonon coupling strength. That correctly expresses the significance of both bare susceptibility and the nesting function below and beyond the Lifshitz transition point. Also, more analyses associated with the polarization of the softened phonon mode at $\textbf q_{CDW}$ explain that instability at this point mainly involves the in-plane displacements of the In atoms and the out-of-plane displacements of the Se atoms at the same time.

Our desired results show that in some hole-doped cases associated with elevated doping levels beyond the Lifshitz transition point ($+0.1, +0.2$, and $+0.3$ e/f.u.), $T_{CDW}$ is much greater than $ T_c$ and consequently, CDW instability prevents access to the superconductive phase, whereas, for other hole doping levels, i.e. doping levels below the Lifshitz transition point ($+0.01$ and $+0.04$ e/f.u.) and very deep hole doping level $+0.4$ e/f.u.,  $T_{CDW}$ is lower than $ T_c$ and a maximum $T_c\sim 75$ K was achieved for low hole doping levels.
In the case of very deep hole doping $+0.4$ and electron doping, rather small $T_c=4$ and $T_c=2$ K, respectively, are obtained. The non-adiabatic phonon effects correctly determining monolayer InSe become dynamically stable for different carrier concentrations at room temperature.

\begin{acknowledgments}
RA was supported by the Australian Research Council Centre of Excellence in Future Low-Energy Electronics Technologies (Project number CE170100039).
\end{acknowledgments}


\nocite{apsrev41Control}
\bibliographystyle{apsrev4-1}
\bibliography{draft1.bib}

\end{document}